\documentclass[journal]{IEEEtran}
 
\usepackage{epsfig}
\usepackage{tabularx}
\usepackage{graphicx,times} 
\usepackage{color}
\usepackage{xspace}
\usepackage{thumbpdf}
\usepackage{listings}
\usepackage{fancyvrb}
\usepackage{verbatim}
\usepackage{hyperref}
\usepackage{booktabs}
\usepackage{colortbl}
\usepackage{relsize}
\usepackage{acronym}
\usepackage{paralist}
\usepackage{flushend}
\usepackage{ifthen}
\usepackage{subfig}
\graphicspath{{artworks/}{graphs/}}
\DeclareGraphicsExtensions{.png,.pdf,.jpeg}
\newcommand{\reffig}[1]{Fig.~\ref{fig:#1}}

\newcommand{\netide}[1][~]{{NetIDE}{#1}}

\newenvironment{packitem}{
\begin{itemize}
  \setlength{\itemsep}{1pt}
  \setlength{\parskip}{0pt}
  \setlength{\parsep}{0pt}
}{\end{itemize}}

\newcommand{\headsup}[1]{\ifthenelse{\boolean{final}}{}{\noindent \textit{\small \color{red} #1}}}
\newcommand{\assignee}[1]{\ifthenelse{\boolean{final}}{}{(#1)}}

\newcommand{\fmer}[1]{\headsup{[ER]~#1}}

\newcommand{\hide}[1]{}

\usepackage{tikz,pgf} %and any other pkgs or tikzlibs your picture needs
\usetikzlibrary{calc,trees,positioning,arrows,chains,shapes.geometric,%
	decorations.pathreplacing,decorations.pathmorphing,shapes,%
	matrix,shapes.symbols}
\usepackage{subfig}
\makeatletter

\begin{document}

% For articles we don't need the environment,
% just the list with acrodef's

\acrodef{amqp}[AMQP]{Advanced Message Queuing Protocol}
\acrodef{api}[API]{Application Programming Interface}
\acrodef{dc}[DC]{data center}%
\acrodef{dmz}[DMZ]{Demilitarised Zone}
\acrodef{ewi}[EWI]{East-West Interface}%
\acrodef{foss}[FOSS]{Free Open-Source Software}
\acrodef{fw}[FW]{firewall}%
\acrodef{ide}[IDE]{Integrated Development Environment}
\acrodef{ietf}[IETF]{Internet Engineering Task Force}%
\acrodef{json}[JSON]{JavaScript Object Notation}
\acrodef{l2sw}[L2SW]{Layer 2 Switch}%
\acrodef{lldp}[LLDP]{Link-layer Discovery Protocol}
\acrodef{nbi}[NBI]{Northbound Interface}
\acrodef{nat}[NAT]{Network Address Translation}%
\acrodef{nemo}[NEMO]{Network Modeling}
\acrodef{netconf}[NETCONF]{Network Configuration}
\acrodef{netide}[NetIDE]{NetIDE}\acused{netide}
\acrodef{nop}[NOP]{No-Operation}%
\acrodef{odl}[ODL]{OpenDaylight}
\acrodef{of}[OF]{OpenFlow}
\acrodef{onf}[ONF]{Open Networking Foundation}
\acrodef{onos}[ONOS]{Open Network Operating System}
\acrodef{qos}[QOS]{Quality of Service}
\acrodef{rtt}[RTT]{Round Trip Time}
\acrodef{sbi}[SBI]{Southbound Interface}
\acrodef{sdn}[SDN]{Software-Defined Networking}
\acrodef{snmp}[SNMP]{Simple Network Management Protocol}
\acrodef{sw}[SW]{Software}
\acrodef{tcam}[TCAM]{Ternary Content Addressable Memory}
\acrodef{vdc}[vDC]{virtual data center}%
\acrodef{xid}[\texttt{XID}]{Transaction Identifier}
\acrodef{xml}[XML]{eXtensible Markup Language}

%%%%%%%%%%%% THIS IS WHERE WE PUT IN THE TITLE AND AUTHORS %%%%%%%%%%%%

\title{Lessons learnt from the NetIDE project:\\ Taking SDN programming to the next level}

%\author{
	%\IEEEauthorblockN{Pedro A. Aranda Guti\'{e}rrez}
	%\IEEEauthorblockA{Telef\'{o}nica I+D, S.A.U., Spain}
	%\and
	%\IEEEauthorblockN{Roberto Doriguzzi-Corin}
	%\IEEEauthorblockA{CREATE-NET, Italy}
	%\and
	%\IEEEauthorblockN{Elisa Rojas}
	%\IEEEauthorblockA{Telcaria Ideas S.L., Spain}
%}

%Paper header
\markboth{Version 1.00}% Version 1.00
{Aranda Guti\'{e}rrez \MakeLowercase{\textit{et al.}}: Lessons learnt from the NetIDE project: Taking SDN programming to the next level}

\author{Pedro A. Aranda Guti\'{e}rrez,
				Roberto Doriguzzi-Corin
        and~Elisa~Rojas% <-this % stops a space
\thanks{Pedro A. Aranda Guti\'{e}rrez is with Universidad Carlos III, Spain. e-mail: paranda@it.uc3m.es}% <-this % stops a space
\thanks{Roberto Doriguzzi-Corin is with FBK CREATE-NET, Italy. e-mail: rdoriguzzi@fbk.eu}% <-this % stops a space
\thanks{Elisa~Rojas is with Telcaria Ideas S.L., Spain. e-mail: elisa.rojas@telcaria.com}% <-this % stops a space
}

\maketitle

\begin{abstract}
  \acl{sdn} promises to overcome \emph{vendor lock-in} by enabling a multi-vendor hardware and software ecosystem in operator networks. However, we observe that this is currently not happening. A framework allowing to compose \acs{sdn} applications combining different frameworks can help revert the trend. In this paper, we analyze the challenges in the current \acs{sdn} landscape and then present the multi-controller \acs{sdn} framework developed by the \acsu{netide} project. Our architecture supports different \acs{sdn} southbound protocols and we have implemented a proof of concept using the \acl{of} protocol, which has given us a greater insight on its shortcomings.
\end{abstract}


\section{Introduction}
\label{section:introduction}

As a flourishing --but young-- paradigm, \acf{sdn} still needs to tackle several challenges. When network operators face adoption of \ac{sdn} in their infrastructures, they are confronted with the need to maximize reusability of use cases and applications. However, currently available network applications are not easily portable from a specific \ac{sdn} deployment to a different one, especially when it implies changing the \ac{sdn} platform they are built upon.
The main reason for this is that \ac{sdn} \acp{nbi} are still not standardized. This causes network applications to depend on the \ac{nbi} implemented by the \ac{sdn} platform they are written for. Which, in turn, causes similar \textit{vendor lock-in} for \ac{sdn} users as they had with network apparel vendors before moving to \ac{sdn}.

\ac{sdn} applications interact with the network in two ways: (i) by generating \ac{sdn} commands for the network infrastructure they control \textit{pro-actively} and (ii) by \textit{reactively} generating \ac{sdn} commands as a response to network events. 
In this paper, we present a novel \ac{sdn} framework developed within the \acsu{netide} project~\cite{netide}, which implements \ac{sdn} application composition for heterogeneous \ac{sdn} controller frameworks. %
We concentrate on the problem space of \textit{reactive} \ac{sdn} applications both in our architectural work and in our implementation.
The Network Engine is complemented by an \acf{ide} that helps \ac{sdn} programmers mix and match the \ac{sdn} platforms that best fit their needs, and benefit from current best practices in programming (e.g. reusing existing and proven code) when developing network applications. %

The whole \ac{netide} framework is available as open source code~\cite{netide_github}. %

\section{Challenges in current SDN landscape}
\label{section:context}

Developing applications for different \ac{sdn} platforms requires different mindsets. 
First of all, the \ac{sdn} programmer has to face the different \textbf{programming languages or \acp{nbi}} that have been used for creating the different \ac{sdn} controller frameworks: e.g. Ryu~\cite{ryu} %\footnote{\url{http://osrg.github.com/ryu/}}
is Python-based while Floodlight~\cite{floodlight}%\footnote{\url{http://www.projectfloodlight.org/floodlight}}
, \acl{odl}~\cite{odl-paper} %\footnote{\url{http://www.opendaylight.org}}
and \acsu{onos}~\cite{onos-paper} %~\cite{onos}
are Java-based. And then, there is the additional burden of the different programming models used by the different frameworks. For example, the \ac{odl} application model builds on an interaction with local data stores that reflect the network nodes controlled by the \ac{sdn} controller. These are modeled using YANG. To write an \ac{odl} application, the first step is to define its YANG model. Then, \ac{odl} internal tools generate the skeletons of the different modules that make up the implementation. In addition, \ac{odl} supports \emph{Maven archetypes} as a means to automate the process.

Staying on the realm of Java-based \ac{sdn} controller frameworks, we have \ac{onos}. It supports \emph{archetypes} too. However, the underlying libraries used to interact with the network elements are quite different: for example, \ac{onos} introduces an intermediate layer called \emph{network intents} which, in essence, is a programmatic abstraction library,  to interact with the network devices without the need for YANG models. The closest \ac{odl} gets to these network intents is MAPLE~\cite{maple}. Other projects labeled as \emph{intent} in \ac{odl} diverge significantly in nature and approach, empowering the user instead of the programmer.

In addition to all this, the \textbf{\ac{ewi}} would be able to grant the synchronization of network applications on different \ac{sdn} platforms. However the implications of the \ac{ewi}
need to be further researched and standardized. In this line, mechanisms to resolve conflicts between applications are also needed. %should be designed, which could be part of the E/WBI or independent from it.
Currently, the most popular \ac{sdn} frameworks (\ac{odl}, \ac{onos}, etc.) implement simple composition mechanisms 
applying priorities for the execution of events, but do not handle conflicts.

Finally, the \ac{onf} %\footnote{\url{http://www.opennetworking.org}}
also includes the \textbf{management functions} in their definition of the \ac{sdn} architecture~\cite{onf-arch-16}. However, many \ac{sdn} platforms ignore them in practice. Including management tools in their networks is thus tedious for operators, who end up implementing their own tailored management solutions; hence increasing the operational cost. 

\begin{table}[!thb]
	\begin{center}
  \begin{tabular}{ m{1.5cm}  m{2.75cm}  m{2.75cm} }
    \hline
    \textbf{Type of NBI} & \textbf{Advantages} & \textbf{Disadvantages} \\ \hline
    Specific programming languages (e.g. Java, Python) & Optimal for application development & Implies rewriting the application even when they share the language (as the API might not be the same)\\ \hline
    REST & Applications from different SDN platforms might share the REST API & Managing every aspect of an application via REST is slower than other APIs (as it is based on a TCP/HTTP connection)\\ \hline
		%GUI \& CLI  \\ \hline
	Intents & High level abstraction, potentially \ac{sdn} framework independent & Still in a very early phase, currently still platform dependent (i.e. not standardized)
    \\ \hline
  \end{tabular}
	\caption{Most common \ac{nbi} for development of SDN applications: pros and cons}
	\label{tab:nbi-options}
	\end{center}
\end{table}
Thus, we can say that despite softwarizing the network, we are not taking advantage of the goodies of software development (i.e. modularization leading to re-usability, etc.).  Table~\ref{tab:nbi-options} shows the different alternatives \acp{nbi} with their pros and cons that a network architect needs to take into account when trying to integrate network applications of different nature in the same \ac{sdn} environment. Since there is no one-fits-all solution, the only alternative seems to be selecting a specific \ac{sdn} environment and redeveloping those applications that are not compatible with it.

\begin{table}[!tb]
	\begin{center}
  \begin{tabular}{ m{2cm}  m{2cm}  m{3cm} }
    \hline
    \textbf{Challenge} & \textbf{Mean to solve it} & \textbf{Why not sufficient} \\ \hline
    Combining heterogeneous SDN applications & Common \ac{nbi} & \ac{nbi} still not standardized. Each SDN platform fight for its own \textit{final} \ac{nbi} that will solve application development\\ \hline
    Combining heterogeneous SDN applications & Common \ac{ewi} & \ac{ewi} still not standardized. Composition and conflict resolution mechanisms still to be discussed in the research community\\ \hline
		Network control and management & Definition of management interfaces & Current solutions based on specific architectures. Most SDN platforms do not even implement it\\ \hline
  \end{tabular}
	\caption{Summary of the challenges in the current SDN landscape}
	\label{tab:summary}
	\end{center}
\end{table}

\ac{sdn} faces two main challenges to avoid \emph{platform lock-in}: we need to be able to reuse and combine \ac{sdn} applications and integrate heterogeneous network management platforms with the \ac{sdn} control plane. Table~\ref{tab:summary} shows current approaches and why they may still not suffice to overcome these challenges.

\section{Related work}
\label{section:related}

Several solutions have been proposed to allow network applications implemented for different \ac{sdn} control platforms to control a common infrastructure. They could be divided into two different groups: online approaches, which try to combine the application modules during run time, and offline approaches, which perform the merging before the execution.

Regarding the \textbf{online} approaches, hypervisors like \emph{FlowVisor}~\cite{flowvisor} and \emph{OpenVirteX}~\cite{openvirtex} split the traffic into ``slices'', which permit the execution of multiple network applications but impede them to cooperate in processing the same traffic as a single network application.

\emph{CoVisor}~\cite{covisor} brings together the following features:
\begin{inparaenum}[(i)]
	\item assembly of multiple controllers,
	\item definition of abstract topologies and
	\item protection against misbehaving controllers.
\end{inparaenum}
In particular, it allows the administrator to combine different client controllers in parallel, in sequence, or in an override/default relationship. However it has several limitations, e.g. it makes the adaptation of an \ac{sdn} operational set-up difficult by forcing the replacement of a running \ac{sdn} controller with a network hypervisor; it can not recognize when an \ac{sdn} application has finished processing a network event, thus potentially leading to network deadlocks.
\emph{FlowBricks}~\cite{flowbricks_concise} is designed to compose client controllers, but \hide{it} only runs on a emulated environment with heavy hacks on the \acl{of} switches and cannot be used over off-the-shelf \hide{\textit{standard}} network hardware.

Approaches based on module orchestration are: \emph{Corybantic}~\cite{corybantic}, which resolves conflicts over specific OpenFlow rules, \emph{Statesman}~\cite{statesman}, which defines three views of the network and only lets application modules propose changes into one of them --later merged by the orchestrator--, and \emph{Athens}~\cite{athens}, which is a compromise between Corybantic and Statesman, allowing more participation in the state that will be sent to the network. However, they all require \ac{sdn} applications to interface with it through a specific \ac{api}, forcing the administrator to modify the applications s/he wants to reuse from different \ac{sdn} environments.

Some other composition options still based on the \ac{nbi} are based on alternative \ac{nbi}. \emph{PGA}~\cite{pga} aims to merge \ac{sdn} applications representing them as graphs and even considering automatic composition. \emph{Redactor}~\cite{redactor} is based on the declarative programming language Prolog and resolves conflicts via a heuristic approach. Other authors leverage CoVisor to provide anomaly-free policy composition~\cite{anomaly-free}.

Finally, FlowConvertor~\cite{flowconvertor} works at \ac{sbi} level, translating policy updates from any origin pipeline to any target pipeline, where one example of pipeline is the OpenFlow switch. However, it does not consider merging different pipelines into a single one.
% ~\cite{wang17} ~\cite{capability-aware}

In the case of the \textbf{offline} approaches, some authors introduce the \emph{Semantics Rule (SR)} concept~\cite{sdn-compiler}, similarly to the Intermediate Representation (IR) for PC compilers, so that SDN application modules are first compiled into SRs (front end), optimized afterwards and finally translated into network low-level rules. Some others propose a \emph{Model-Driven Networking (MDN)} framework~\cite{sdn-modeling} where modules are translated --or directly written-- into a Domain-Specific Modeling Language (DSML), allowing easy merging and verification, and finally generating the corresponding code for a targeted SDN platform.

\section{\acs{netide} Network Engine architecture}
\label{section:architecture}

As mentioned in the introduction, the \ac{netide} project has developed an SDN platform with a companion IDE~\cite{netide_netsoft16} that provides a true cross-platform development and deployment experience. The \ac{sdn} platform, called Network Engine, is based on a three-tier approach, with a layer of client controllers executing applications and a layer of server controllers driving the network elements. The Network Engine's controllers are orchestrated by a core layer that includes composition and conflict resolution mechanisms. In this Section we provide a high-level overview of the NetIDE Network Engine architecture. A detailed description is available on~\cite{netide_cnsm}.

\begin{figure}[!h]
  \centering
	\includegraphics[width=0.44\textwidth]{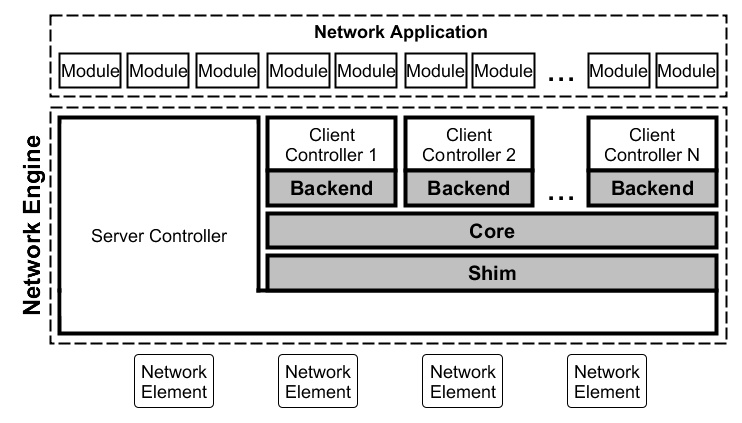}
  %\vspace*{-.25em}
  \caption{Network Engine architecture.}
  \label{fig:netide-architecture}
  %\vspace*{-1em}
\end{figure}

The proposed Network Engine (\reffig{netide-architecture}) combines unmodified \ac{sdn} applications running on multiple client controllers (called \textit{Modules} in the figure), organizing them to cooperate with modules running on top of the network-facing controller (represented by the \textit{Server Controller} in the figure). 
In this regard, a \textit{Network Application} is understoods as a set of \ac{sdn} software modules, possibly written for different \ac{sdn} controller platforms, which are orchestrated to cooperate on controlling the same network. %
These network applications behave as single entities that compute \ac{sdn} combining the rules produced by the modules it is comprised of with custom-defined semantics.

The most challenging aspect of the Network Engine is integrating client and server controllers. A first idea is to connect the \ac{sbi} for the client controllers to the \ac{nbi} of the server controllers. But as these interfaces do not normally match, adaptation is necessary. To maximize reuse, we use separate adaptors for the \ac{sbi}, called \textit{Backend}, and for the \ac{nbi}, called \textit{Shim}. 
An intermediate layer, or \emph{Core}, communicates with the Shims and Backends. It implements the \ac{sdn} controller framework agnostic functions, thereby making the implementation of both Shims and Backends for new \ac{sdn} frameworks light-weight.

%!TEX root=netide.tex
\subsection{The Core} \label{section:intermediate_layer}

The \textbf{Core} is a platform-independent component that %interfaces with Backends and Shim and
orchestrates the execution of individual modules, that are potentially spread across multiple controllers. It controls all messages exchanged between application modules and the network. In this sense, the operations of the Core can be divided into three categories:
\begin{inparaenum}[(i)]
\item handling the asynchronous events from the network, such as new flows, port status, flows removed,
\item composing the configuration messages generated by the application modules and checking them for conflicts, and
\item pairing \textit{read-state} request messages issued by the application modules %(e.g. \texttt{FEATURE\_REQUEST}, \texttt{GET\_CONFIG} in \acl{of})
  and the corresponding replies from the network.
\end{inparaenum}

The Core intercepts network events from the server controller and distributes them to the client controllers based on a \textit{Composition Specification} file that defines which modules are used in the composition and the flow of execution between them. The Core supports two execution semantics: \textit{Sequential} and \textit{Parallel} (akin to the Sequential and Parallel operators in~\cite{frenetic,pyretic}).
\begin{packitem}
\item{Sequential} execution invokes modules in the sequence \hide{\fmer{in the sequence? / following the sequence?}} defined by the composition specification. The first module is fed the original input event; % (e.g. the original \texttt{PACKET\_IN} event, in an OpenFlow environment);
  each subsequent module uses the original input modified with the actions returned by previous modules in the chain.
\item{Parallel} execution invokes modules in parallel using the same input for all. 
\end{packitem}
The Core implements \textit{policies} to merge the actions returned by the modules into a consistent set of actions that can be installed in %down into
the network to handle the traffic. Policies are used to determine how conflicting outputs are handled; options include checking
%when the same
for conflicts in a specific field of an action. % contains contradictory instructions, such as \texttt{drop} and \texttt{forward}.
Standard policies are \textit{discard} conflicts by dropping conflicting results; %ing results by discarding them),
\textit{ignore} conflicts by installing \emph{all} results without further processing;  %resolving the conflicts)
and \textit{prioritize} and resolve conflicts by picking the result with the highest priority out of the set of conflicting results.

%Operation (iii) is performed by a message \textit{routing} component. With \textit{read-state} messages we refer to requests issued by application modules to collect information from the network (e.g. \texttt{FEATURE\_REQUEST}, \texttt{GET\_CONFIG} in \acl{of}). The Core ensures that the responses generated by the network for such requests arrive to the right module, i.e. by the module that made the request. \acl{of} uses the \ac{xid} to ease the pairing. However, in the context of the Network Engine, \acp{xid} are not sufficient, as different modules may use the same value, effectively making the reply/request pairing impossible. The Core avoids duplicated \ac{xid} values by replacing the identifiers it finds in the requests with unique values. 

\subsection{Shim and Backend}\label{section:backend-shim}

The \textbf{Shim} and \textbf{Backend} are %controller
platform-specific components that integrate existing controller frameworks into the \ac{netide} architecture.

\begin{figure}[h!]
	\centering
	\includegraphics[width=1\linewidth]{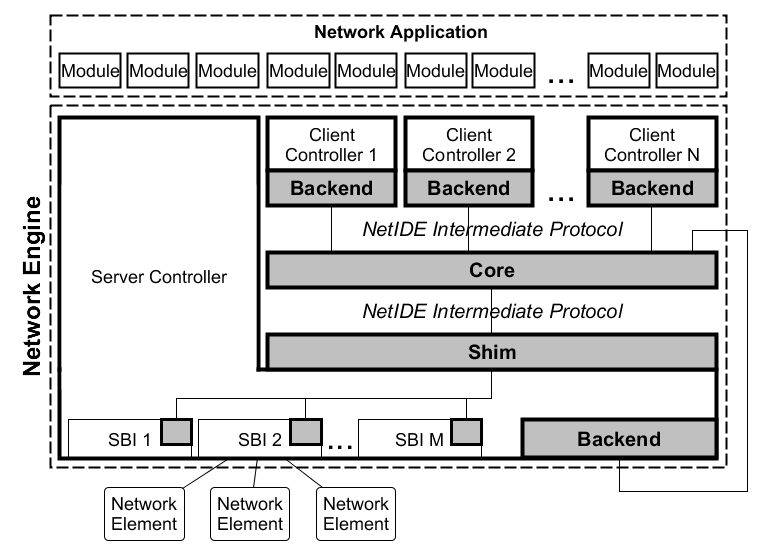}
	\caption{Detailed architecture of the Network Engine.}
	\vspace{-1.5em}
	\label{fig:detailed-architecture}
\end{figure}

The {Shim} translates the \ac{nbi} of the server controller to the \netide \ac{api}, exposing it to the Network Engine.
As shown in \reffig{detailed-architecture}, it overrides the server controller's processing logic in the \ac{sbi} and routes all messages from the network to itself.

The {Backend} is an additional \ac{sbi} for the client controller that interacts
with the underlying layers of the Network Engine. At boot-time, the Backend registers the application modules running on the client controller to the Core, which, in turn, assigns a specific identifier to each of them. 

At runtime, Backends use the module identifier 
to identify the module sending the message. 
On the other direction, the Core uses the identifier based on pre-defined policies to indicate which module handles an event. The Backend steers the event distribution inside the client controller, ensuring events are sent to the correct modules.

\reffig{detailed-architecture} shows a detailed view of our architecture. We include in it the mechanism we use to integrate modules written for the server controller framework into our applications.
Specifically, we place a Backend in the server controller to steer the message flow for the modules composing Network Application that run on the server controller, exactly like for any other module running on a client controller. In this case, the modules can only interact with the Backend, since the other \acp{sbi} are hidden to them by the Shim, as explained above in this section.   

\subsection{NetIDE Intermediate Protocol} \label{section:intermediate_protocol}

The \netide Intermediate protocol implements the following functions:
\begin{inparaenum}[(i)]
	\item to carry management messages between the Network Engine's layers (Core, Shim and Backend); e.g., to exchange information on the supported \ac{sbi} protocols, to provide unique identifiers for application modules, implement the \textit{fence} mechanism,
	\item to carry event and action messages between Shim, Core, and Backend, properly demultiplexing such messages to the right module based on identifiers, and
	\item to encapsulate messages from different \acp{sbi} to a common format handled by the Core.
\end{inparaenum}

It includes fields to identify network elements for \ac{sbi} messages, application modules running on the client controllers and transactions (i.e. groups of related network events and commands as described above).

\subsection{Fencing}
\label{section:fence}

The way different \ac{sdn} frameworks, and particularly \acf{of}, work is that \ac{sdn} modules receive network events and \emph{optionally} produce network commands in response.
This implies that modules may quit silently, without producing any tangible response.
%However, in some cases, the module just quits silently producing no output.
A prerequisite to implement application composition 
is to know when all modules of an application have finished processing a network event. %This is the prerequisite to trigger the composition.
Otherwise, there is a risk that the Core performs composition operations too early, i.e. when some modules are still processing the event, or that it freezes waiting for a response which will never arrive. %This is often called \emph{``run-to-completion''} problem.\\
We introduced \textit{fences}, i.e. end-of-execution markers, to tackle this problem and require that Backends monitor the execution flow within the client controllers to this avail. 
The Core supports interleaved communication with the application modules to improve the performance of the Engine. In order to preserve the semantic of the network policies installed by the modules, the Core ensures that the composed output goes back to the network consistently with the time ordering of events.

\section{Practical scenario}
\label{section:usecases}

\begin{figure}[!h]
  %\centering \subfloat[]{\includegraphics[width=0.23\textwidth,natwidth=568,natheight=704]{uc1-datacentre-engine1a}\label{fig:uc1-datacentre1a}}
	\centering \subfloat[]{\includegraphics[width=0.23\textwidth]{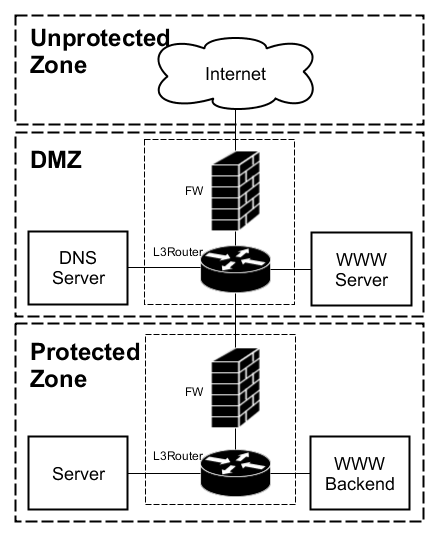}\label{fig:uc1-datacentre1a}}
  %\hspace{0.5em} \subfloat[]{\includegraphics[width=0.227\textwidth,natwidth=569,natheight=713]{uc1-datacentre-engine1b}\label{fig:uc1-datacentre1b}}
	\hspace{0.5em} \subfloat[]{\includegraphics[width=0.227\textwidth]{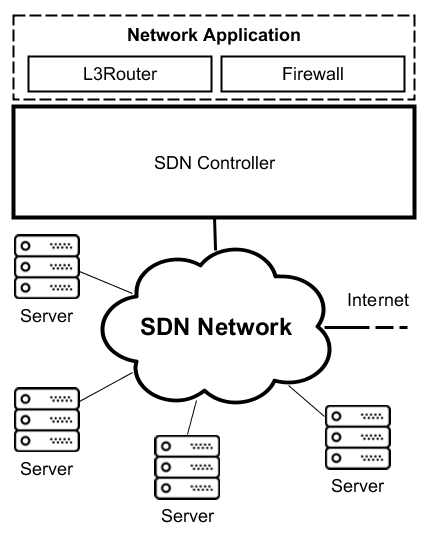}\label{fig:uc1-datacentre1b}}
  \caption{\label{fig:uc1-datacentre1}An \ac{sdn}-driven \ac{dc}.}
\end{figure}

One of the scenarios that has driven the research and implementation work of the project is the \ac{sdn}-driven \acl{dc}, where we explore the possibility of implementing the networking components of a \ac{vdc} service as \ac{sdn} applications. 
We consider a typical \ac{vdc} offering including components like an unprotected zone connected to the Internet, a \ac{dmz} and an interior zone. 
The logical topology of the \ac{vdc} (shown in \reffig{uc1-datacentre1a}) is implemented using \ac{sdn} applications written for a specific \ac{sdn} controller platform (\reffig{uc1-datacentre1b}) to provide firewall and Layer 3 network services, and stitching between the different virtual machines that implement the individual components of this \ac{vdc} offering such as Web and DNS services.

We observe that there is a logical grouping into zones (exterior zone, \ac{dmz} and interior zone) that can be used as predefined blocks in \ac{vdc} implementations \hide{\fmer{to flexibly implement vDCs? / to provide flexibility in vDC implementations?}:  \fmer{IEEE recommends using 'for example' here instead of e.g.} } for example, a provider can offer micro-\acp{vdc} that consist of a \ac{dmz} only, \acp{vdc} with more than one isolated \ac{dmz} for different organizations within a company, etc. In turn, each of these building blocks uses atomic building blocks like the \acl{fw}, router, etc. This structure calls for the use of \textit{patterns}, component reuse and other software development techniques. 

The operator may require to activate new network services on the operational \ac{vdc} to improve, for instance, the performance and the security of the network. Thus, she/he may want to spread the users’ requests over different servers via a Load Balancer, or to hide part of the internal network behind a single IP address using a \ac{nat} service.% \remove{(\reffig{uc1-datacentre2a})}.

\begin{figure}[!h]
  \centering
%\subfloat[]{\includegraphics[width=0.22\textwidth,natwidth=608,natheight=839]{uc1-datacentre-engine2a}\label{fig:uc1-datacentre2a}}
	\subfloat[]{\includegraphics[width=0.22\textwidth]{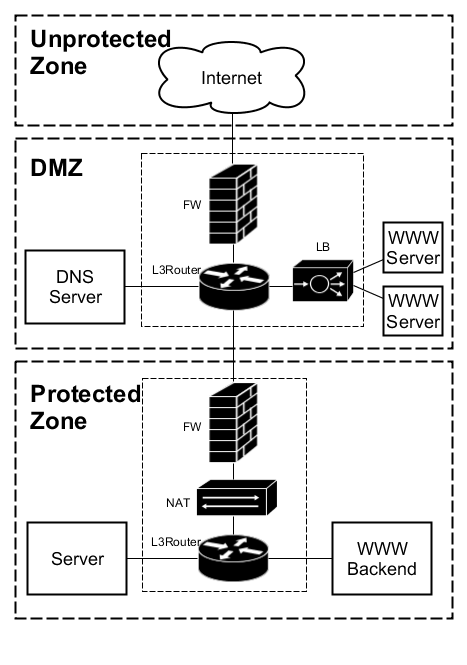}\label{fig:uc1-datacentre2a}}
  \hspace{0.5em}
  %\subfloat[]{\includegraphics[width=0.235\textwidth,natwidth=683,natheight=876]{uc1-datacentre-engine2b}\label{fig:uc1-datacentre2b}}
	\subfloat[]{\includegraphics[width=0.235\textwidth]{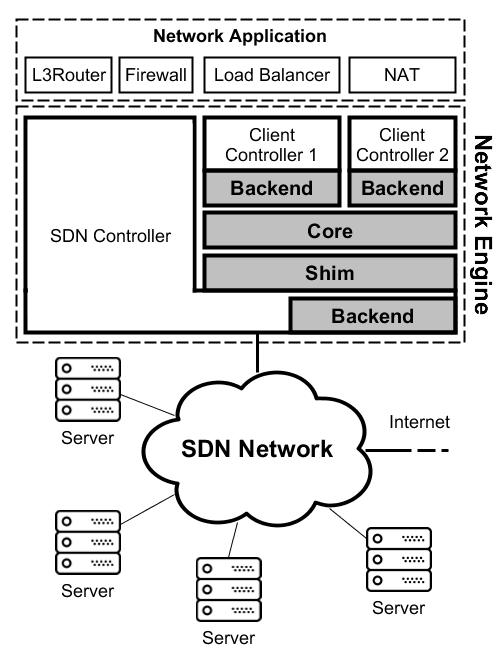}\label{fig:uc1-datacentre2b}}
  \caption{\label{fig:uc1-datacentre2}Evolved \ac{sdn}-driven \ac{dc}.}
\end{figure}

With \ac{netide}, the operator enhances the existing \ac{vdc} module shown in \reffig{uc1-datacentre1a} with the router and firewall modules, which executed are as-is on different \ac{sdn} controller frameworks (as shown in \reffig{uc1-datacentre2b}), instead of porting them to --or writing them from scratch for-- the platform he/she initially chose.

\section{Lessons learnt and Shortcomings of \acl{of}}
\label{section:shortcomings}
The architecture we present in this paper implements the composition of \ac{sdn} applications. We have concentrated on \acl{of} for our implementation because many, diverse controller frameworks for OpenFlow are available as \ac{foss}. Although the experience we have gathered shows the shortcomings of the approach taken by the OpenFlow community, many of them can be generalized to other state-of-the-art \ac{sdn} environments.

On the one hand, most frameworks have no way of telling from the outside when a \emph{controller has  completely consumed} a network event. This resulted in run-to-completion problems in our composition core, until we introduced the fencing mechanism. 
Partially coupled with this, we also had to struggle with the fact that OpenFlow also lacks \emph{clearly defined \acl{nop} semantics}, i.e. how to interpret the situation when a controller consumes a network event and produces no output.

The separation of control and data plane into different entities, which communicate through a standardized interface is the most valuable contribution of OpenFlow to the development of \ac{sdn}.
However, \emph{timing matters} in \ac{sdn} applications; when writing composed applications, we have experienced situations where independently running modules may introduce transient network state that then influences the way the network responds to network events in an uncontrolled matter. 
In the OpenFlow model, applications are not aware of the \emph{actual} network state. When they need to be, they have to reproduce it internally in the application and it is then where spurious interactions can have a significant (and negative) impact on the network behavior.

\ac{netide} represents a consistent approach to parallel composition of \ac{sdn} applications. Sequential composition, that is using the response of a module as the input for another one, %or implementing $C(ev_i) = M_2(M_1(ev_i))$ (in a mathematical representation)
is an issue which deserves further study~\cite{Schwabe:ARNW2016}. The output of a module is a combination of network commands and generated packets and transforming that into valid network events for the subsequent modules in a sequential composition chain is not evident:
%due to their disjoint natures.
first of all, there is no one-to-one transformation from input to output events of \ac{sdn} modules because their very natures are not reconcilable and, secondly, there is the need of a snapshot of the network state to try any approximate transformation.

\section{Conclusion} \label{section:conclusion}

In order to introduce software development paradigms in \acl{sdn}, we have developed the \ac{netide} architecture. It provides an \ac{ide} to lower the entry barrier to \ac{sdn} and a Network Engine that allows composing \ac{sdn} application using pre-existing ones as building blocks.

Our Network Engine concept is \ac{sbi}-independent and we have implemented it for \acl{of} as a proof-of-concept. In this process, we had to face all the shortcomings for the OpenFlow paradigm, including insufficiently defined semantics in some operations, incompatible semantics of events, coexistence of \ac{sdn} application models that can not be composed, etc. We have overcome some of these limitations with our architecture and have contributed both architectural concepts and a fully functional development and runtime architecture to the \ac{sdn} community.

From our experience, we strongly feel that facing all these shortcomings and developing an evolved architectural model for \ac{sdn} would result in a more robust yet flexible architecture for future software-defined networks, which truly overcomes vendor or platform \emph{lock-in} risks.

%{\small
\section*{Acknowledgments}
\noindent  The work presented in this paper has been partially sponsored by the European Union through the FP7 project NetIDE, grant agreement 619543.
%}

\bibliographystyle{IEEEtran} 
\bibliography{netide}

\begin{thebibliography}{10}
\providecommand{\url}[1]{#1}
\csname url@samestyle\endcsname
\providecommand{\newblock}{\relax}
\providecommand{\bibinfo}[2]{#2}
\providecommand{\BIBentrySTDinterwordspacing}{\spaceskip=0pt\relax}
\providecommand{\BIBentryALTinterwordstretchfactor}{4}
\providecommand{\BIBentryALTinterwordspacing}{\spaceskip=\fontdimen2\font plus
\BIBentryALTinterwordstretchfactor\fontdimen3\font minus
  \fontdimen4\font\relax}
\providecommand{\BIBforeignlanguage}[2]{{%
\expandafter\ifx\csname l@#1\endcsname\relax
\typeout{** WARNING: IEEEtran.bst: No hyphenation pattern has been}%
\typeout{** loaded for the language `#1'. Using the pattern for}%
\typeout{** the default language instead.}%
\else
\language=\csname l@#1\endcsname
\fi
#2}}
\providecommand{\BIBdecl}{\relax}
\BIBdecl

\bibitem{netide}
``{NetIDE Project},'' \url{http://www.netide.eu}.

\bibitem{netide_github}
``{Network Engine source code},'' \url{https://github.com/fp7-netide/Engine}.

\bibitem{ryu}
``{Ryu SDN framework},'' \url{http://osrg.github.com/ryu/}.

\bibitem{floodlight}
\BIBentryALTinterwordspacing
{Project Floodlight}. [Online]. Available:
  \url{http://www.projectfloodlight.org/floodlight/}
\BIBentrySTDinterwordspacing

\bibitem{odl-paper}
J.~Medved, R.~Varga, A.~Tkacik, and K.~Gray, ``{OpenDaylight: Towards a
  Model-Driven SDN Controller architecture},'' in \emph{Proceeding of IEEE
  International Symposium on a World of Wireless, Mobile and Multimedia
  Networks 2014}, June 2014, pp. 1--6.

\bibitem{onos-paper}
P.~Berde, M.~Gerola, J.~Hart, Y.~Higuchi, M.~Kobayashi, T.~Koide, B.~Lantz,
  B.~O'Connor, P.~Radoslavov, W.~Snow, and G.~Parulkar, ``{ONOS: Towards an
  Open, Distributed SDN OS},'' in \emph{{Proceedings of the Third Workshop on
  Hot Topics in Software Defined Networking}}, ser. HotSDN '14, New York, NY,
  USA, 2014, pp. 1--6.

\bibitem{maple}
``{SDN Programming using High Level Programming Abstractions},''
  \url{https://wiki.opendaylight.org/view/SDN_Programming_using_High_Level_Programming_Abstractions}.

\bibitem{onf-arch-16}
\BIBentryALTinterwordspacing
ONF. (2016) {SDN architecture - Issue 1.1}. [Online]. Available:
  \url{https://www.opennetworking.org/images/stories/downloads/sdn-resources/technical-reports/TR-521_SDN_Architecture_issue_1.1.pdf}
\BIBentrySTDinterwordspacing

\bibitem{flowvisor}
R.~Sherwood, M.~Chan, A.~Covington, G.~Gibb, M.~Flajslik, N.~Handigol, T.-Y.
  Huang, P.~Kazemian, M.~Kobayashi, J.~Naous, S.~Seetharaman, D.~Underhill,
  T.~Yabe, K.-K. Yap, Y.~Yiakoumis, H.~Zeng, G.~Appenzeller, R.~Johari,
  N.~McKeown, and G.~Parulkar, ``{Carving Research Slices out of Your
  Production Networks with OpenFlow},'' \emph{SIGCOMM Comput. Commun. Rev.},
  vol.~40, no.~1, pp. 129--130, Jan. 2010.

\bibitem{openvirtex}
A.~Al-Shabibi, M.~De~Leenheer, M.~Gerola, A.~Koshibe, G.~Parulkar,
  E.~Salvadori, and B.~Snow, ``{OpenVirteX: Make Your Virtual SDNs
  Programmable},'' in \emph{Proceedings of the Third Workshop on Hot Topics in
  Software Defined Networking}, ser. HotSDN '14, 2014, pp. 25--30.

\bibitem{covisor}
X.~Jin, J.~Gossels, J.~Rexford, and D.~Walker, ``{CoVisor: A Compositional
  Hypervisor for Software-defined Networks},'' in \emph{Proceedings of the 12th
  USENIX Conference on Networked Systems Design and Implementation}, ser.
  NSDI'15, 2015, pp. 87--101.

\bibitem{flowbricks_concise}
A.~Dixit, K.~Kogan, and P.~Eugster, ``{Composing Heterogeneous {SDN}
  Controllers with Flowbricks},'' in \emph{22nd {IEEE} International Conference
  on Network Protocols, {ICNP} 2014, Raleigh, NC, USA, October 21-24, 2014},
  2014, pp. 287--292.

\bibitem{corybantic}
J.~C. Mogul, A.~AuYoung, S.~Banerjee, L.~Popa, J.~Lee, J.~Mudigonda, P.~Sharma,
  and Y.~Turner, ``{Corybantic: Towards the Modular Composition of SDN Control
  Programs},'' in \emph{Proceedings of the Twelfth ACM Workshop on Hot Topics
  in Networks}, ser. HotNets-XII, 2013, pp. 1:1--1:7.

\bibitem{statesman}
P.~Sun, R.~Mahajan, J.~Rexford, L.~Yuan, M.~Zhang, and A.~Arefin, ``{A
  Network-state Management Service},'' in \emph{Proceedings of the 2014 ACM
  Conference on SIGCOMM}, 2014.

\bibitem{athens}
A.~AuYoung, Y.~Ma, S.~Banerjee, J.~Lee, P.~Sharma, Y.~Turner, C.~Liang, and
  J.~C. Mogul, ``{Democratic Resolution of Resource Conflicts Between SDN
  Control Programs},'' in \emph{Proceedings of the 10th ACM International on
  Conference on Emerging Networking Experiments and Technologies}, 2014.

\bibitem{pga}
C.~Prakash, J.~Lee, Y.~Turner, J.-M. Kang, A.~Akella, S.~Banerjee, C.~Clark,
  Y.~Ma, P.~Sharma, and Y.~Zhang, ``{PGA: Using Graphs to Express and
  Automatically Reconcile Network Policies},'' in \emph{Proceedings of the 2015
  ACM Conference on Special Interest Group on Data Communication}, ser. SIGCOMM
  '15, 2015, pp. 29--42.

\bibitem{redactor}
W.~Wang, W.~He, and J.~Su, ``{Redactor: Reconcile network control with
  declarative control programs In SDN},'' in \emph{2016 IEEE 24th International
  Conference on Network Protocols (ICNP)}, Nov 2016, pp. 1--10.

\bibitem{anomaly-free}
M.~Rezvani, A.~Ignjatovic, M.~Pagnucco, and S.~Jha, ``Anomaly-free policy
  composition in software-defined networks,'' in \emph{2016 IFIP Networking
  Conference (IFIP Networking) and Workshops}, May 2016, pp. 28--36.

\bibitem{flowconvertor}
H.~Pan, G.~Xie, Z.~Li, P.~He, and L.~Mathy, ``{FlowConvertor: Enabling
  Portability of SDN Applications},'' \emph{IEEE INFOCOM 2017}, 2017.

\bibitem{sdn-compiler}
\BIBentryALTinterwordspacing
H.~Li, C.~Hu, P.~Zhang, and L.~Xie, ``{Modular SDN Compiler Design with
  Intermediate Representation},'' in \emph{Proceedings of the 2016 Conference
  on ACM SIGCOMM 2016 Conference}, ser. SIGCOMM '16.\hskip 1em plus 0.5em minus
  0.4em\relax New York, NY, USA: ACM, 2016, pp. 587--588. [Online]. Available:
  \url{http://doi.acm.org/10.1145/2934872.2959061}
\BIBentrySTDinterwordspacing

\bibitem{sdn-modeling}
F.~A. Lopes, L.~Lima, M.~Santos, R.~Fidalgo, and S.~Fernandes, ``{High-level
  modeling and application validation for SDN},'' in \emph{NOMS 2016 - 2016
  IEEE/IFIP Network Operations and Management Symposium}, April 2016, pp.
  197--205.

\bibitem{netide_netsoft16}
P.~A.~A. Guti\'{e}rrez, E.~Rojas, A.~Schwabe, C.~Stritzke, R.~Doriguzzi-Corin,
  A.~Leckey, G.~Petralia, A.~Marsico, K.~Phemius, and S.~Tamurejo, ``{NetIDE:
  All-in-one framework for next generation, composed SDN applications},'' in
  \emph{2016 IEEE NetSoft Conference and Workshops (NetSoft)}, 2016.

\bibitem{netide_cnsm}
R.~Doriguzzi-Corin, P.~A.~A. Gutierrez, E.~Rojas, H.~Karl, and E.~Salvadori,
  ``Reusability of software-defined networking applications: A runtime,
  multi-controller approach,'' in \emph{2016 12th International Conference on
  Network and Service Management (CNSM)}, Oct 2016, pp. 209--215.

\bibitem{frenetic}
N.~Foster, R.~Harrison, M.~J. Freedman, C.~Monsanto, J.~Rexford, A.~Story, and
  D.~Walker, ``{Frenetic: A Network Programming Language},'' in
  \emph{Proceedings of the 16th ACM SIGPLAN International Conference on
  Functional Programming}, ser. ICFP '11, 2011, pp. 279--291.

\bibitem{pyretic}
C.~Monsanto, J.~Reich, N.~Foster, J.~Rexford, and D.~Walker, ``{Composing
  Software-defined Networks},'' in \emph{Proceedings of the 10th USENIX
  Conference on Networked Systems Design and Implementation}, ser. nsdi'13,
  2013, pp. 1--14.

\bibitem{Schwabe:ARNW2016}
\BIBentryALTinterwordspacing
A.~Schwabe, P.~A. {Aranda Guti{\'e}rrez}, and H.~Karl, ``{Composition of SDN
  Applications: Options/Challenges for Real Implementations},'' in
  \emph{Proceedings of the 2016 Applied Networking Research Workshop}, ser.
  ANRW '16.\hskip 1em plus 0.5em minus 0.4em\relax New York, NY, USA: ACM,
  2016, pp. 26--31. [Online]. Available:
  \url{http://doi.acm.org/10.1145/2959424.2959436}
\BIBentrySTDinterwordspacing

\end{thebibliography}

\label{last-page}
\end{document}